# CARBON-BASED NANOCOMPOSITES: PROCESSING, ELECTRONIC PROPERTIES, AND APPLICATIONS

**Manab Mallik \*, Mainak Saha**

Department of Metallurgical and Materials Engineering, National Institute of Technology Durgapur, India
\*E-mail of Corresponding Author: manab.mallik@mme.nitdgp.ac.in

**Abstract:**
The last two decades, in particular, have witnessed a large volume of research revolving around structure-property correlation in carbon-based nanocomposites synthesized by several methods. In the simplest of terms, the electronic properties of these nanomaterials, which form the present context of discussion, vary mainly as a function of **three** parameters, out of which **two** are process parameters (viz. (i) the kind of reinforcement and (ii) method of synthesis), and **one** is a structure-dependent parameter. **The structure-dependent parameter is** highly influenced by the **two** process parameters and plays a vital role in determining the ionic and electronic transport phenomenon in these materials. **In other words,** the interaction between electrons and the equilibrium 0-D (point) defects, along with different types of interfaces, plays a crucial role in the understanding of electronic properties, apart from the physical and chemical properties **of** these materials. The present chapter **provides** a brief overview of state-of-the-art research and detailed discussions on some recent developments in understanding the electronic properties of some conventional carbon-based nanocomposites (synthesized by several different techniques) on the structure-property correlation in these materials using the **three** parameters mentioned above. **F**inally, some of the significant challenges in this field **have been addressed** from both industrial and fundamental viewpoints.

## 1. Introduction

The sole and highly-explored notion of developing a "hybrid" material with integrated properties of **two** or more components, since the last **two** decades, has paved the way for the design of several "complex" materials, beyond the conventional concept of material development based on structure-property correlation **[1]**. Nanocomposites are one such class of materials, which at present find applications in a large number of sectors, ranging from nanoelectronics to energy storage industries due to their exceptional electronic, mechanical and chemical properties. In other words, these materials have entirely revolutionized the world of "functional materials" and hence, may be considered as materials of the 21$^{st}$ century with the exploration of newer avenues on a highly frequent basis.

Nanocomposites are defined as composite or multiphase materials in which as a minimum one phase has at least one dimension in nanoscale ($10^{-9}$ m) range [2]. Nanocomposites have extraordinary physical, mechanical, and chemical properties, which make these materials appropriate alternatives to conventional materials and composites.

Among different nanocomposites, carbon-based nanocomposites have witnessed a large volume of attention in the last two decades due to their structure-dependent electronic properties, low density, and large specific surface area [1,3]. Carbon-**based** nanocomposites **contain** different carbonaceous materials (graphite/ nanotube/ $C_{60}$) as matrix **with** various reinforced phases such as metal oxides/ sulfides/ nitrides etc.

Besides, in recent times, due to adsorbates' high charge transfer capacity, carbon Nanotube (CNT) based nanostructured composites have proven to be promising materials for application in electrochemical supercapacitors, gas and biological sensors, electromagnetic absorbers, and photovoltaic cells [3]. This has arisen due to limited charge sensitive conductance in Single-walled (SW) CNTs, like in graphite [4, 5]. Hence, the novel strategies of combining CNTs with transition metals like Ru, Pt, etc. [6, 7] and metallic oxides or conducting polymers have been attempted [8], and a higher specific capacitance value compared to CNTs in supercapacitor applications have been obtained. However, the exceptionally high price of these materials limits their application in commercial supercapacitors [8]. Besides, the need to cope up with the ever-increasing demands for higher energy density and power output of rechargeable Li-ion batteries (LIBs) [9] has necessitated the usage of CNT based nanocomposites both as cathode and anode materials for such application. Recent work has also highlighted the use of graphite (with a suitable binder) as anode material for Potassium Ion Batteries (PIBs) to improve energy efficiency than that of LIBs and Sodium-Ion Batteries (SIBs) [10].

Moreover, in the recent decade, the "correlative microscopy" **methodology** [11-16] has been used extensively used as a tool to understand a number of properties in materials especially **used** for different structural applications. One of the main reasons **as to why the**"correlative microscopy" **methodology** has not been used widely to study structure-property correlation in C-based nanocomposites, may be attributed to the challenges involved in sample preparation. **Moreover, there are other challenges associated with the characterisation of** a light element, **such** as C and understanding the complexity of the atomic structures in nanocomposites. This has also been the reason **behind a limited understanding o**f the structure-dependent parameter. **Thus,** most of the literature published in this field are able to address the structure-property correlation as a function of **only** process parameters **(and not structure-dependent parameter)** in C- based nanocomposites. As already discussed, the present chapter limits its premises in discussing only the electrical properties of nanocomposites from the state-of-the-art on research and finally, some of the major problems, requiring a high level of systematic analysis, in order to create a new paradigm in the field of research on functional materials. However, the authors do not claim the present review to completely address all the aspects of understanding the electrical properties of nanocomposites due to the countless number of papers already published in this field.

## 2. Why is carbon so interesting?

Amongst all the elements in the periodic table, carbon turns out to be extremely interesting, not only due to its ability to form several versatile isotopes, playing a vital role in the development of human civilization but also because of the ability to give rise to several completely different properties, on account of being bonded differently with other atoms and molecules. In addition to the costly diamond and graphite, various other carbon forms include nanotubes, fullerene molecules, graphene, etc. (Fig. 1) [17–22]. Different forms of carbon allotropes are shown in Fig. 1 [21-**24**].

In the well-known diamond cubic crystal structure with each carbon atom bonded to four other carbon atoms by strong $sp^3$ covalent bonds, the absence of free (mobile) electrons makes diamond a hard yet highly electrically insulating material. graphite, on the other hand, is formed by the stacking of 2-D hexagonal flat sheets of $sp^2$-bonded carbon atoms with the sheets weakly bonded by **V**an der Waals forces. This bond interestingly leads to the high mobility of C between the sheets and primarily accounts for the high electronic conductivity of graphite. Besides, the narrow bandgap between conduction and valence bands in graphite also accounts for excellent electrical conductivity in graphite. Before the onset of carbon nanotubes (CNTs) [25, 26] and then graphene, the most investigated artificially synthesized allotrope of C was fullerene $C_{60}$ molecule with an icosahedral structure (having **twenty** hexagons and **twelve** pentagons) in a truncated manner. Charge doping may also be employed as a method to induce superconductivity in the semiconducting $C_{60}$ molecule [17, 18, 25].

For defining CNTs, a chiral vector ($c=na_1+ma_2$, with $a_1$ and $a_2$ being the unit vectors of the 2-D hexagonal lattice of graphene and n and m being integers) is used. Based on the number of graphene layers present, CNTs may be broadly classified into two types: Single-walled CNTs (SWCNTs) and Multi-walled CNTs (MWCNT) built from one or few single layers of graphene, with an interlayer distance of nearly 0.34 nm. Graphene, on the other hand, exhibits extraordinary electronic properties with electrons behaving as Dirac fermions and showing tremendously high mobility together with holes [9, 20]. Besides, the band structure of graphene with the fully occupied valence band with the empty conduction band (both touching at **six** points) renders a zero-gap semiconductive nature to graphene [20].

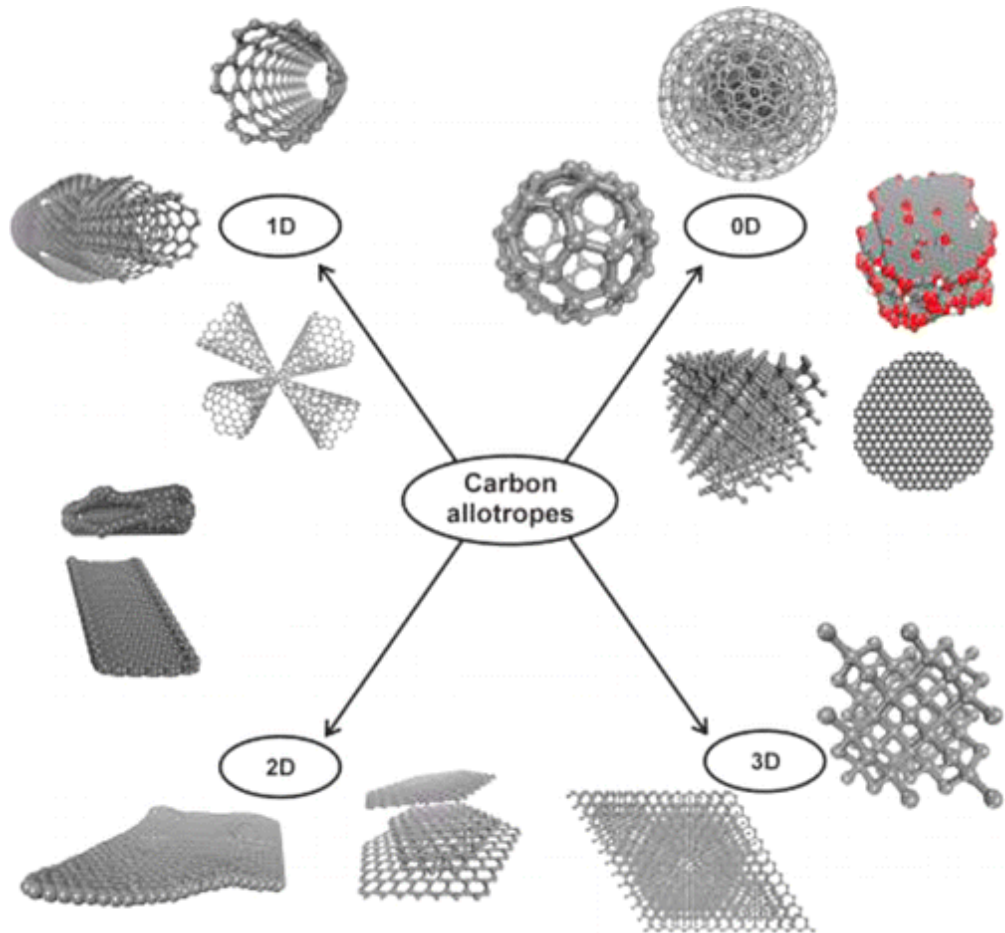

**Fig. 1** Structures of different forms of carbon allotropes (**Reproduced with permission from [21]**)

**2.1 Carbon nanotube (CNT) based nanocomposites**

Apart from the wide usage of CNTs owing to their properties [**22-**25], there are certain exceptional features rendering CNTs as an ideal supplement for a large group of materials, including metallic materials, polymers, and ceramics. The present state of research in CNTs is mainly aimed at synthesizing CNT-based composites as an alternative to conventional materials [25]. However, compared to CNTs, graphene-based materials exhibit numerous advantages: a unique combination of mass production coupled with low manufacturing [25] and excellent electronic conductivity [26]. In the context of electronic properties, it is highly interesting to discuss the intensively studied $RuO_2$ reinforced CNT used to design supercapacitor devices, having a much higher cost and possessing toxic characteristics among inorganic supercapacitor electrode materials [3]. The use of CNT-based metal oxides/conducting polymers as electrochemical supercapacitor electrode materials have not only resulted in an extremely high specific capacitance (~180 F/g [4,5]) but has also led to higher electronic conductivity when compared to that in activated carbon [3-6]. CNTs may be synthesized by several methods ranging from arc discharge to different Chemical Vapour Deposition (CVD) techniques, and even a significant industrial waste (especially from coke ovens in integrated steel plants) such as coal tar may also be used as the starting material for the same [27]. The radius of a CNT, typically varying between 100 nm to 20 cm, strongly depends on synthesis parameters [3], which strongly influence the electronic properties of the material. CNTs are reported to be covered by a half of a fullerene-like molecule [7].

The present state of research on CNT-based nanocomposites mostly revolves around the detailed understanding of supercapacitor behavior and the stability of CNT based metal oxides and CNT based polymers through the engineering of the structure-based parameter which involves composition, grain size etc. [27], in addition to enhancing the electronic, mechanical, and thermal properties of these materials [27- 32]. Kavita et al. [28] have studied the influence of poly (vinyl butyral), and structural modification (based on acid functionalized MWCNT treatment) on the thermomechanical properties of the novolac epoxy nanocomposites and reported an increment of nearly ~15 °C at the peak degradation temperature in comparison to the unmodified novolac epoxy. In recent times, a double-layer capacitor with fullerene-activated carbon composite electrodes has been reported by Okajima et al. [32] to possess an extremely high capacitance even on a 1 wt% ultrasonically treated electrode (loaded with $C_{60}$). The first attempt to synthesize a novel Pt-modified $[PW_{11}NiO_{39}]^{5-}$@reduced graphene oxide (rGO) and Pt-modified $[PW_{11}NiO_{39}]^{5-}$@multiwall carbon nanotube (MWCNT) composites, has been reported by Ensafi et al. [31]. Pyrrole-treated functionalized SWCNT exhibits excellent electrochemical performance, which includes a high capacitance and power density [33]. The plasma surface treatment of MWCNTs with $NH_3$, leading to an enhancement of total surface area and wettability of MWCNTs and thus, an enhancement in capacitance, has been reported by Yoon et al. [34]. CNTs/conducting polymer composites, on the other hand, have been synthesized by several techniques, among which the most common are the in-situ polymerisation (both chemical and electrochemical polymerisation) of monomers [35, 36].

### 2.1.1 Polymer CNT based composites

Owing to an excellent combination of thermal, optical, and electrical properties [25], as already discussed in the earlier section, CNTs turn out to be potential candidates in various newly-emerged fields, which commonly include nanotechnology and biosensors. Polymer reinforced CNT nanocomposites possess an excellent combination of mechanical properties and electrical conductivity for biosensor applications [35]. In the context of supercapacitor devices, the performance is largely determined by MWCNT content in the composite electrode [37]. As a result, polypyrrole (PPy)/MWCNT composites may be employed as potential candidates for supercapacitors with high capacitance and a long life per cycle [8, 37].

The composite electrodes composed of conducting polymers reinforced in CNTs results in an enhancement of mechanical strength along with thermal and electronic conductivity [38]. PANI/ MWCNT composite has been reported to possess a specific capacitance combined with a large retention of capacitance even after a large number of cycles [39]. Besides, a number of conducting polymers, mainly polyaniline, polythiophenes polypyrrole, have also been reported to possess high capacitance [40]. This has rendered CNT based conducting polymer nanocomposites to be promising materials for energy storage application in supercapacitors [41- 44], and many amperometric biosensors [41, 42]. Raman spectra of i-PANI/MWCNTs and r-PANI/MWCNTs (PANI- Polyaniline) composites are represented in Fig. 2(a), wherein the Raman peaks (characteristic of PANI) were observed to undergo a significant change with temperature. XRD patterns of PANI/MWCNT composites are shown in Fig. 2(b) with PANI molecular conformations, schematically illustrated in Fig. 2(c) as a function of temperature [43]. The electronic properties of PANI, poly (vinylidene fluoride) (PVDF) and MWCNTs, was studied by Blaszczyk-Lezak et al. [41] wherein, the surface functionalization of MWCNT was performed in a concentrated mixture of sulfuric and nitric acid at 90 °C for 24 h.

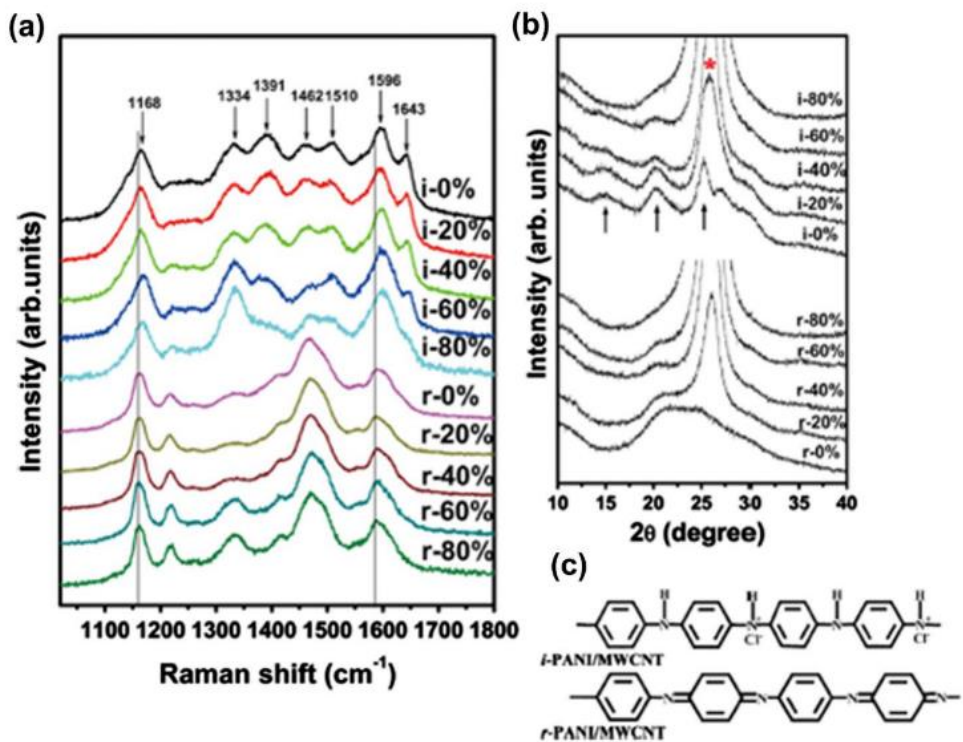

**Fig. 2** Representation of (a) Raman spectra and (b) XRD patterns of PANI reinforced MWCNT composites with variation in weight fraction of MWCNTs (* represents CNT peaks) (c) schematic illustration of PANI molecular conformations at different temperatures of synthesis **(Reproduced with permission from [43])**.

## 2.1.2 Activated Carbon (AC)/CNT based nanocomposites

As reported by Navarro-Flores et al. [45], activated carbon (AC)/CNTs nanocomposite electrodes on being tested in an organic electrolyte (1.5 M NEt$_4$BF$_4$ in acetonitrile) render a number of interesting results, the most significant of which is a reasonably good compromise between energy and power density even for a CNT content of 15 wt.% [45]. **Besides, a high cell series resistance of ESR (~ 0.6 Ωcm$^{-2}$) and high capacitance (~88 F/g) have also been reported from this work [45]**. The possibility of preparing AC/CNT nanocomposite based electrodes for supercapacitor devices using electrophoretic deposition (EPD) technique has been extensively explored by Huq et al. [46] wherein, it was observed that the as-prepared AC/CNT nanocomposite electrode shows excellent capacitance retention (of nearly 85%) even after a cyclic stability test for prolonged time (~ 11,000 cycles [46]). Besides, AC has also been used extensively as an electrode material in Electric Double Layer Capacitors (EDLCs) for a prolonged period owing to a remarkable combination of high capacitance, long cycle life, and most importantly, low cost of manufacturing [47]. **Qiu et al. [48] have reported that activated hollow carbon fibers (ACHFs) containing CNTs and Ni nanoparticles (CNTs-Ni-ACHFs) may be synthesized using various techniques such as thermal reduction and Chemical Vapour Deposition (CVD)**. However, to date, activated carbon (AC) remains as the most commonly used absorbent, primarily, owing to its high surface area [49-51].

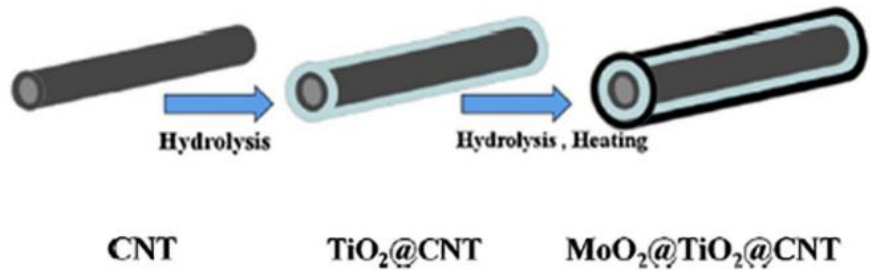

**Fig. 3** Schematic of synthesis processes for MoO$_2$@TiO$_2$@CNT nanocomposites (with sandwich-like structure) (**Reproduced with permission from [52]**).

## 2.1.3 Metal oxide/CNT nanocomposites

Most of the studies on electronic properties of CNT based on the RuO$_2$/TiO$_2$ nanocomposites have considered the potential window (-0.4 to +0.4 V) [53]. Alam et al. [54] have synthesized BaMg$_{0.5}$Co$_{0.5}$TiFe$_{10}$O$_{19}$/MWCNT nanocomposites with variation in the amounts of MWCNTs (0, 4, 8, and 12 vol%) and demonstrated that ~8% vol. of MWCNTs in the synthesized MXCNT based nanocomposite possesses the best microwave absorption capacity. Besides, A novel method to increase the interfacial bond strength by developing a coating of magnesium oxide (MgO) nanoparticles on the CNT substrate has been reported by Yuan et al. [55]. Metal oxides often find suitable candidates for pseudo-capacitive electrode materials for supercapacitor devices due to their high-energy density [55], unlike carbonaceous materials. Titanium dioxide ($TiO_2$) is presently a vital anode material for rechargeable Li-ion batteries due to an excellent combination of high cycle life coupled with high safety and low cost. Yuan et al. [52] have reported the synthesis of $MoO_2@TiO_2@CNT$ nanocomposites (sandwich structured) from CNTs through a **two**-stage process involving (i). Hydrolysis of CNT to form $TiO_2@CNT$ and finally, (ii) Hydrolysis and heating(in an inert $Ar/H_2$ atmosphere) of $TiO_2@CNT$ to form $MoO_2@TiO_2@CNT$, as schematically illustrated in Fig.3.

### 2.1.4 Carbon fibers (CF)/CNT based nanocomposites

Based on the study by Islam et al. [56], CNTs, covalently bonded with CF through an ester linkage in the absence of catalysts or coupling agents, may be employed to acquire the combined advantages of enhanced interfacial shear and impact strength [8]. The reason lies in the requirement for a higher level of tensile stress to be able to detach the CNTs from CF [57]. **There are two methods used at present to reinforce CF with CNTs using physical adsorption technique (based on weak Van der Waals interaction) limit the reinforcing effect [58]**. CF/CNT based nanocomposites have found an extensive application in the context of lightweight automotive and aerospace components with high fuel efficiency- "The present need of the hour."
Wang et al. [58] have reported an improvement in the interfacial shear strength of CFs using graphene oxide (GO) as reinforcement with CNT. CNTs have numerous advantages as conducting wires as compared to copper wires owing to their size and quantum effects [59]. Tamrakar et al. [60] deposited poly-ethylenimine (PEI) functionalized multi-wall carbon nanotube (CNT) using EPD technique onto the S-2 glass fiber surface and reported that EPD method provides the required thickness of CNT coating, thereby facilitating control of interfacial resistivity between fiber and matrix.

### 2.1.5 CNT/metal nitride-based nanocomposites

Jiang et al. [61] have reported the electronic properties of CNT/metal nitride-based nanocomposites with a remarkable property that this material retains the electrochemical stability of the metal nitride, even in presence of strongly corrosive electrolytes. The presence of CNTs enhances the electronic conductivity of the CNT/TiN composites through Percolation effect (of CNTs) [61- 63]. Based on the published literature in this field, the mobility of carry conduction electrons through the lattice to numerous scattering sites, ranging from point defects to different kinds of 2-D interfaces and numerous 3-D volume defects such as pores etc. play a vital role in influencing the electronic properties of TiN [64]. CNTs, likewise, provide an excellent conducting path [61, 65], especially if these are located(or more likely segregated)  at the grain boundaries (GBs) of the composite. The presence of CNTs at the GBs aids carrier transportability and improves the electronic conductivity of the composite [63]. This is one of the aspects of Materials Science research where the "less well known" structure-dependent parameter starts influencing the electronic properties of the C-Based nanocomposites. CNTs have also been reported to serve as an electrically conducting bridge to connect the domains of TiN, primarily due to the higher charge mobility of the CNT–TiN composite compared to that of TiN [63]. The other reported reason is the presence of strong interfacial cohesion between CNTs and the matrix [63] due to the presence of TiN nanoparticles along the CNT wall, leading to a highly efficient transfer of electrons from the matrix to the CNT. The presence of ~12 vol% CNTs in CNT– TiN nanocomposite has been reported to exhibit nearly~ 45% increase in electrical conductivity compared to TiN [64, 65].

## 2.2 Vanadium nitride (VN)/graphene (G) composite

The practical implementation of rechargeable lithium-sulfur batteries, till 2017 had been impeded by several issues, among which the shuttle effect leading to a rapid loss of capacity combined with a low **c**oulombic efficiency **were** the most significant.**, T**he first report on the development of conductive porous vanadium nitride (VN) nanoribbon/graphene (G) composite came from Sun et al. [66]. Based on both experimental ad theoretical results, vanadium nitride/graphene composite has been reported by Sun et al. [66] to provide a strong driving force for rapid conversion to polysulfides. Owing to the high electronic conductivity of VN, the composite cathode has been reported by Sun et al. [66] to exhibit lower polarization coupled with faster kinetics of redox reaction as compared to that in a reduced graphene oxide (rGO) cathode, showing excellent cycling performances [Fig. 4]. Based on this report [66], the initial capacity reaches 1,471 mAhg$^{-1}$, and the capacity after 100 cycles is 1,252 mAhg$^{-1}$ at 0.2 C, with a loss of ~ 15%, thereby offering the potential for use in high energy lithium-sulfur batteries.

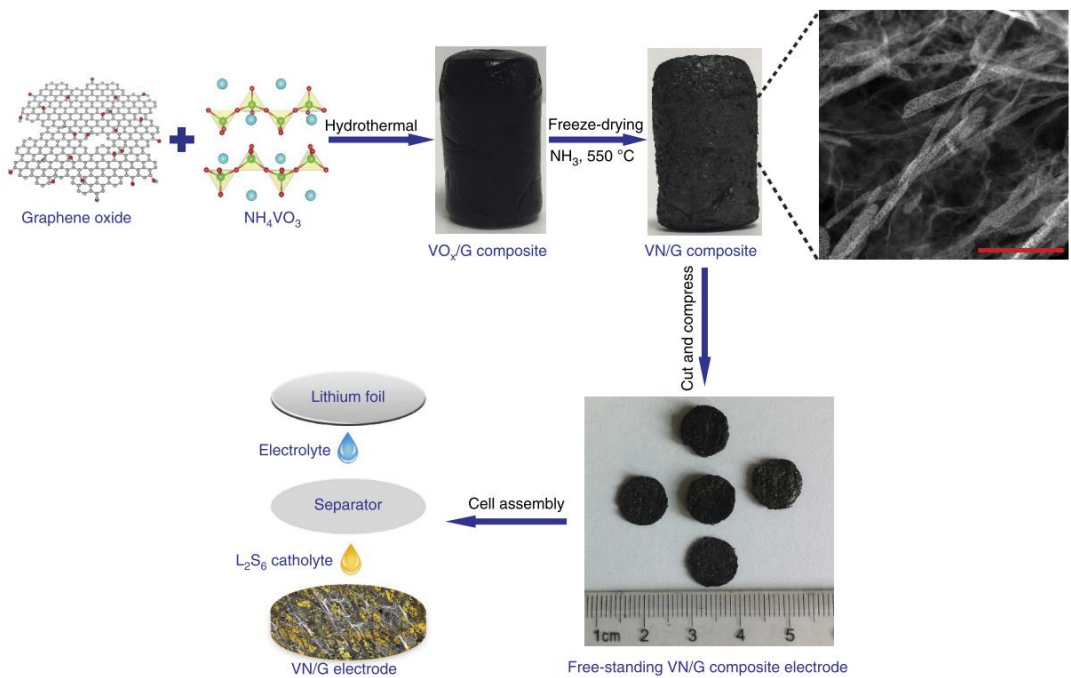

**Fig. 4** The schematic view illustrates the synthesis of a porous VN/G composite electrode and the cell assembly with corresponding optical images of the fabricated material. A scale bar of 500nm has been used **(Reproduced with permission from [66])**.

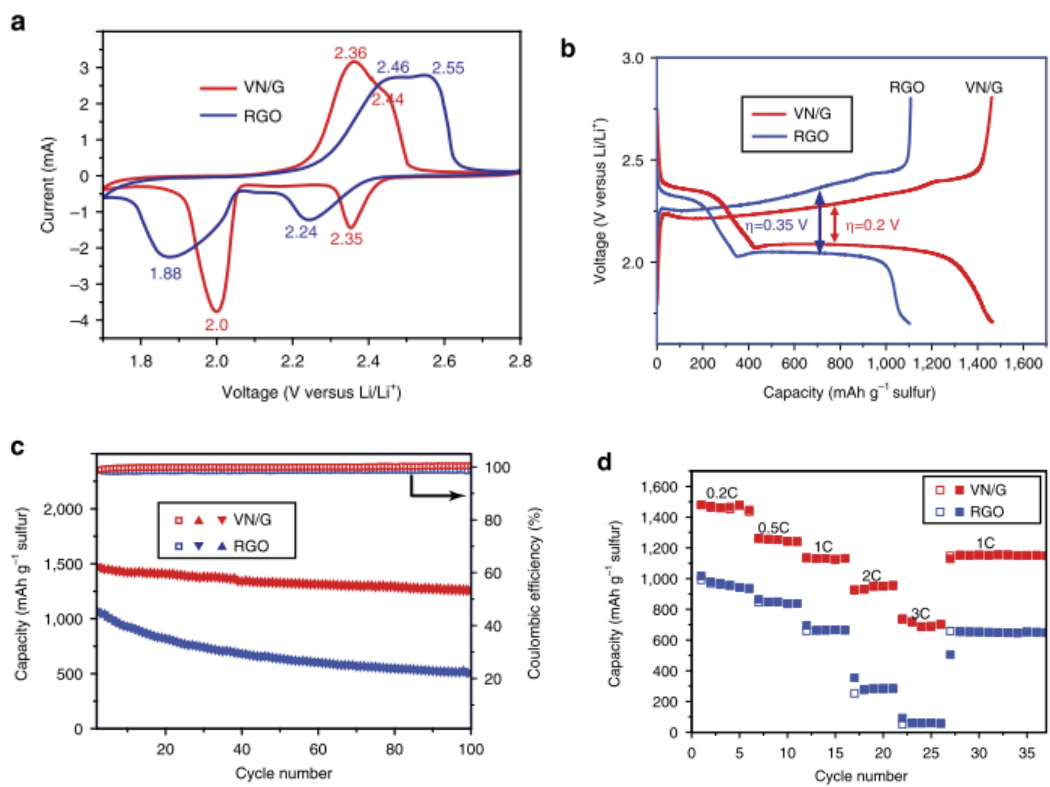

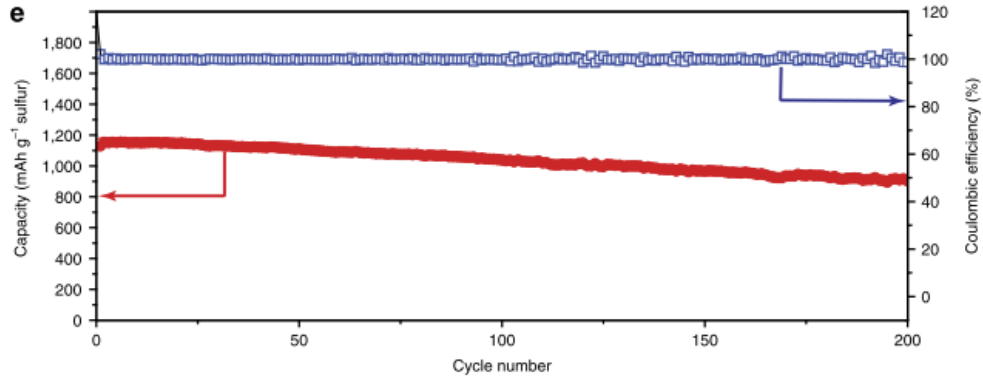

**Fig. 5** Plots illustrating the electrochemical performances of VN/G and rGO cathodes. (a) CV profiles of the VN/G and rGO cathodes (scan rate of 0.1mVs$^{-1}$) between 1.7 to 2.8 V. (b) Galvanostatic charge-discharge profiles of the VN/G and rGO cathodes (at 0.2 C). (c) Cycling performance and Charge (Coulombic) efficiency of the VN/G and rGO cathodes (at 0.2 C for 100 cycles). (d) Rate performance of the VN/G and rGO cathodes at different current densities. (e) Cycling stability of the VN/G cathode (at 1 C for 200 cycles) **(Reproduced with permission from [66])**.

Based on Sun et al. [66], the VN/G nanocomposite, prepared using a 2-stage Hydrothermal and Freeze-drying technique, schematically illustrated in Fig. 4, combines the benefits of both graphene and VN. Fig. 5 shows the XRD pattern and TG-DSC curve for VN/G nanocomposite. Electron and ion transportation, along with electrolyte absorption, is facilitated by the 3D free-standing structure of the graphene (G) network. Besides, the VN/G electrode has also been reported to possess excellent specific capacitance with a much higher as compared to the rGO electrodes. This pathbreaking work [66] has also opened up a new avenue of research wherein even metal nitrides other than VN may be used as potential cathodes to explore energy storage capacity for applications in Li-S batteries.

## 3. Nanocrystalline and amorphous Carbon films (CNx)

Hydrogenated or non-hydrogenated C films, often termed as diamond-like carbon (DLC) films [3], may be either amorphous or nanocrystalline with mixed states of sp$^3$ and sp$^2$ bonding. Those with predominantly sp$^2$ state of bonding are known as graphite-like amorphous Carbon (GAC) films. To be extremely specific, the bonding and antibonding electronic states due to the presence of π bonds (of sp$^2$ carbon sites) determine the electronic properties and many physical properties of these materials [4]. In the context of electronic properties, they are highly attractive due to a wide range of DC electrical conductivity values at ambient temperature (ranging from ~10 to 10$^1$ S/m [4]). A thermally activated process dominates the temperature dependence of electronic conductivity in these materials (with **σ = σ₀ exp(−ΔE/kT): σ: intrinsic conductivity [4], σ₀: pre-exponential factor [4], ΔE: energy barrier [4] and T: absolute temperature [4]**) based on hopping mechanism [4]. Among the existing theoretical models, the variable range hopping (VRH) model [8] is the most commonly used for studying the temperature dependence of electronic conductivity in these materials. For instance, the electronic conductivity of activated carbon (AC) has been reported to change from 1-dimensional (1D) to 3-dimensional (3D) during the graphitization process, depending on the thermal treatment temperature [4]. Besides, at temperatures below 1000 °C under high-pressure conditions, VRH is the primary conduction mechanism, as reported by Zhao et al. [67]. In 1D VRH, the electrical resistivity ρ follows an exponential relationship with temperature T (**ρ = ρ₀ exp(T₀/T)$^x$: ρ: electrical resistivity at temperature T [4], ρ₀: electrical resistivity at infinite temperature [4], T₀: characteristic temperature [4]**) with x = 1/2 for 1-D VRH [5]) with $T_0$ related to localization length ξ for the wave function [6]. Even for large 2D (with x=1/3) or 3D (x=1/4) nanoparticles, VRH conduction mechanisms have been reported, indicating the change in electronic conduction mechanism from 1D to 2D or 3D [4]. The insulator-metal transition has been reported between non-graphitized and almost-graphitized regions. The material exhibits semi-metallic behaviour in the near-graphitized region and a nearly linear electrical resistivity-temperature (ρ-T) relationship. At higher sintering temperatures (1200–1600°C), the electrical resistivity of the graphitized activated C exhibits a power-law relation with temperature (**ρ=A+BT$^{3/2}$: A and B: coefficients of temperature [4]**) with C behaving as a non-Fermi liquid [67].

The concentration of localized states by Hydrogen in hydrogenated DLC films has been extensively reviewed by Staryga et al. [68], where the effects of nitrogen doping on the electronic transport characteristics have also been studied in amorphous $CN_x$ films. These films owing to the localization of sp² hybridized state and interestingly, significant changes in the electrical properties have been observed in amorphous-$CN_x$ (a-$CN_x$) films as a function of the N concentration. It has been correlated to bond strength between N atoms and sp² and sp³ sites of C [69]. The dynamics of hopping transport in a-$CN_x$ have also been studied in detail using AC electrical spectroscopy measurements [70–72].

## 3.1 Diamond-like carbon (DLC) and graphite-like amorphous (GAC) based nanocomposites

One of the most researched topics in the context of electronic applications in DLC and GAC films is their capability to be used in electron field emission devices [4] and as cold-cathodes in field emission displays [73, 74]. Incorporating boron is an effective method to enhance the oxidation resistance of various carbon-based nanocomposites, thereby avoiding the major drawback of surface oxidation in these materials. These materials' electronic conduction is generally reported in terms of Mott VRH for localized electronic states near the Fermi level [73, 74]. Porosity, coupled with the highly inert nature of DLC and GAC, render them ideal candidates as matrix for neutral or electron donor nanoparticle species to prepare hybrid materials for catalytic applications [74]. In devices such as rechargeable Li-ion batteries (LIBs) meant for energy storage with high energy density, $V_2O_5$/C composites have also been observed to improve high-rate performances [75].

## 3.2 Hard nanocomposite coatings with amorphous carbon

In the context of amorphous C (a-C) based nanocomposite thin films and hard nanocrystalline C-based coatings, there is a lack of understanding of C atoms' influence on the electronic properties of these materials. For instance, C segregation at GBs is generally expected to lead to the formation of localized phases at GBs, as recently reported by Meiners et al. [76], which depends on GB's structure, maybe either conducting or insulating [4]. Although, DC electrical resistivity measurements as a function of temperature may be extensively used as an experimental tool to investigate the structural evolution (involving grain size of both the matrix phase and also of the phase present at GBs) of the composite films. However, the concept of "Correlative Microscopy," discussed in the Introductory section of the review has not been widely employed in these materials, till date, to understand the presence of GB phase in these materials, although an extensive review on the influence of structured interfaces on the electronic performance of composite materials, has been published by Mishnaevsky Jr. [77]. Fig. 6 shows the variation in electrical resistivity due to structural modifications in WC/a-C nanocomposites. The electrical resistivity vs. temperature curves suggest that electrons' scattering against 2-D GBs and O-D point defects turn out to be extremely significant. In such nanocomposite materials, the experimental results have been reported to be easily validated by the GB scattering model [78]. Based on calculations using this particular model (Fig. 6), the deduced grain sizes have been reported to be highly consistent with crystallite sizes estimated from different structural characterization techniques such as XRD and TEM [4]. Secondly, the calculated elastic scattering free path ($l_e$) values have been reported to be in line with a high density of point defects in the films [4]. Thirdly, grain size (D) has been reported to be the main parameter controlling the electronic transport phenomenon of these films. Based on the present understanding of the vanishing of GBs in quasi-amorphous materials and that the scattering of electrons is largely dominated by the high density of 0-D (point) defects, mainly vacancies as reported by Sanjinés et al. [4].

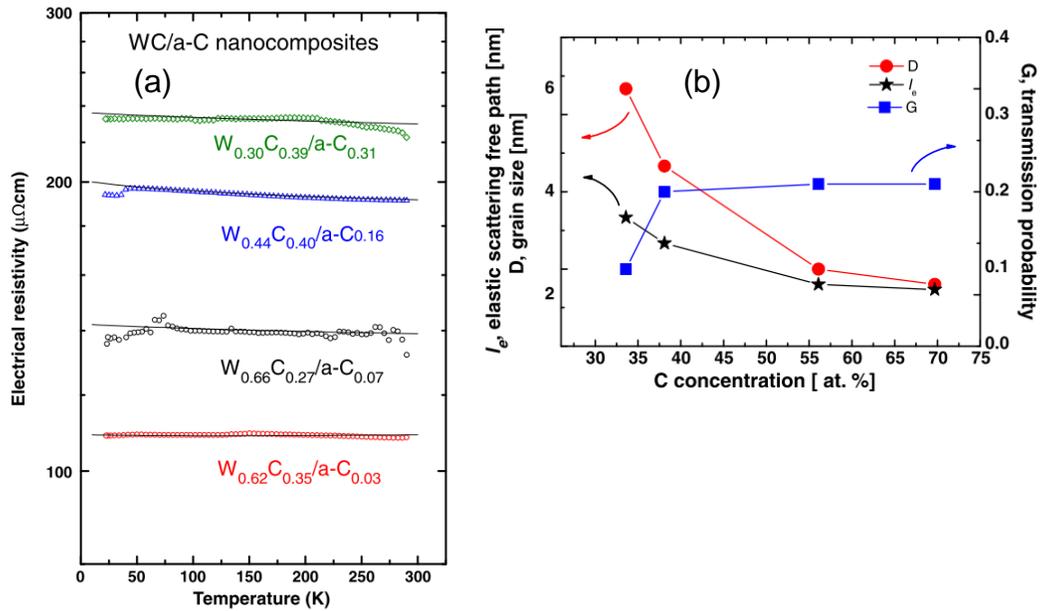

**Fig. 6** (a) Plot illustrating DC electrical resistivity vs. temperature of a-C/WC nanocomposites. The solid lines (in the figure) represent the best fit with the GB model. (b) Elastic scattering free path ($l_e$), grain size (D), and electron transmission probability (G) deduced from the model as a function of atomic concentration of C (**Reproduced with permission from [4]**).

## 4. Some of the major limitations and future challenges in this field

### 4.1 From an industrial viewpoint

In the context of Supercapacitors, presently, power limitation is a significant issue that depends on the type of applied electrolyte, electrode type, and several other parameters such as evaluated constant current, to name a few, and much further investigation on the electrodes and their charge-storage mechanisms are needed. However, the present state of research in supercapacitors' energy-storage capacity is focused primarily on the understanding of charge storage mechanisms (in sub-micropores) based on material design [80, 81]. Electrical energy is stored in the bulk structure of electrodes, whereas the electrodes' surface area opposes this particular phenomenon of "charge storage inside the battery electrodes" [8]. At present, both SWCNTs and MWCNTs have been extensively used in supercapacitors owing to their high specific surface area, low electrical resistance, high cyclic stability, and low mass density [8, 79, 80]. Besides, several investigations have been carried out on exploring the charge-storage capacities of C-based materials such as activated carbon, porous carbon, CNTs, graphene, etc. [79, 80].

Davies et al. [81] reported that oxides and nitrides of transition metals, as well as conducting polymers, have the potential to improve the electrochemical performance of supercapacitors. These materials are considered as pseudocapacitive materials. Also, as reported by Ates et al. [8], EDLCs exhibit remarkably good cycling stability but lower specific capacitance, whereas, on the contrary, pseudocapacitive materials have low stability during cycling process owing to the reactions (faradaic type).

Hence, in recent times, there has been a considerable focus on utilizing a hybrid material based on the combined advantages of both EDLCs and Pseudocapacitative materials, through combining rapid capacitive charging with high energy density. Although, at present, the attempts aimed at incorporating C to redox materials to enhance capacitive ability appears advantageous; however, a detailed understanding of the influence of pore size distribution and process parameters, mainly involving synthesis techniques on electrochemical performance, is currently limited. It is only through a proper investigation of some fundamental aspects (in microstructural level) in these materials that supercapacitor devices may be made to render a high energy density (similar to the currently rechargeable Li-ion batteries (LIBs)) to the Electrochemical Capacitors (ECs), thereby leading to a significant impact on the future application of high power energy storage devices. There are numerous technological challenges to the use **of** CNT-based nanocomposites as electrode materials for LIBs, Sodium-Ion Batteries (SIBs) [9] and the recently developed Potassium Ion Batteries (PIBs) [9]. to the **limiting factors mainly include** poor adhesion between at CNT/nanoparticle interface **and** problems of uniform dispersion of nanoparticles in CNTs [8].

As discussed by Liu et al. [9], other significant challenges include the development of environment-friendly and industrial fabrication techniques for these materials for application in the battery systems. In recent times, **graphene** was used as an alternative to other carbon allotropes, including CNTs, for many applications, especially with the development of various routes for synthesis [82-85]. Similar to **g**raphene, **g**raphene oxide (GO) also exhibits an interesting combination of different properties, including enhanced electrical conductivity and mechanical strength with a high level of non-toxicity and large electrochemical stability. These combinations of properties make both graphene and graphene oxide as suitable candidates for electrode materials for LIBs and supercapacitors [86-88]. Although graphene may be used as a replacement for CNT-based nanocomposites in the electrodes of rechargeable LIBs, SIBs, and PIBs, in the context of energy storage applications, **h**owever, the significant challenge viz. the development of an economic fabrication technique for large-scale production of these materials still needs to be overcome [87].

Besides, as reported by Guo at al. [89] and Chang et al. [90], the globally increasing demands for energy storage applications have paved an avenue for many investigations aiming at industrial applications of Graphene-based nanocomposites, through modifications in mainly the process parameters, associated with their fabrication methods and subsequent understanding of structure-property correlation based on such modifications. It is worth mentioning that the enormous potential of graphene-based nanocomposites in energy storage applications has been recognized by a large number of cutting-edge research [91-104] in a number of areas such as photoelectrochemical and photovoltaic devices, LIBs, supercapacitors, etc. In particular, graphene nanocomposite-based LIBs and supercapacitors presently turn out to be highly promising materials for energy storage applications.

**However, there remain several unexplored avenues and a countless number of challenges to be addressed before one may implement graphene-based nanocomposites in real industrial applications of graphene-based nanocomposites** exist. The most important of them all is the mass production of high and uniform quality graphene nanocomposites. Secondly, photoelectrochemical and photovoltaic efficiencies in devices based on graphene nanocomposites are still reported to be relatively low even in most prototype devices [90-94]. Hence, as the fundamental questions remain addressed despite considerable research from different groups is getting published almost regularly. The idea of aiming for industrial applications with these materials is nearly close to a myth. However, the challenges associated, especially with the fabrication and both long and short term industrial applications of graphene nanocomposites, make them one of the most exciting materials for energy storage and energy conversion-related studies and, hopefully, may be achieved in the near future.

**As discussed in an extensive review on interfacial engineering of Li-ion batteries (solid-state) (LIBs) [105], the major drawbacks associated with research in the field of LIBs have been due to the fact that those have been directed only towards obtaining highly efficient ionic conductors in the form of Solid-state electrodes (SSEs) rather than designing batteries for practical applications**. This is further reflected in the fact that the LIBs, used today, still have a number of limitations, such as low rate capability coupled with limited energy and power density and low cycle life [105]. The limitations are mostly in terms of properties of the solid-solid interfaces (between SSEs and other cell components such as cathode and anode) in the LIBs. However, recently, there has been a gradual shift from seeking better SSEs to designing Solid-state Batteries (SSBs) with high interface stability and interfacial resistance through the design of an artificial buffer layer for the modification of buffer layer and (/or) surface of the electrode. For a rational design of the synthetic buffer layer, a detailed understanding of the challenges associated with the solid-solid interfaces viz's electronic properties. SSE/cathode and anode/SSE interfaces are highly essential. The challenges related to the interface between SSE and cathode/anode are mainly due to:

i) High resistance (at the interface) and consequent uneven current distribution during the operation of the SSB due to limited physical contact at the interface between SSE and cathode/anode leading to poor rate performance and subsequent formation of Li dendrites [105].

ii) Miserable cycle life combined with low energy and power density associated with volume and stress changes of the electrode materials during charge-discharge cycles may quickly deteriorate physical contact [105].

iii) Formation of a thick passivated layer hindering the diffusion of $Li^+$ across the electrode/SSE interfaces due to chemical or/and electrochemical side reactions between the electrode and SSE [105] and

iv) Formation of a space-charge layer at the interface due to direct contact between two ionic conductors (e.g., the active material in the cathode and the ceramic electrolyte) with different chemical potentials of $Li^+$ ions, thereby deteriorating both rate and cycling performance [105].

The addition of a soft polymer-based buffer layer has also been reported to be an effective way to partially address the above challenges [106]. However, due to low electronic conductivity at room temperature and poor mechanical strength against Li dendrites (in particular), polymer-based buffer layers are unable to completely overcome the challenges to balance the conductivity, safety, and thickness [105, 106].

In the context of interfacial stability with cathodes, the development of a coating between the cathode materials with an inorganic artificial buffer layer has been reported to be a promising approach for improving the stability of the cathode/SSE interface, mainly for sulfide-type SSEs [105]. However, volumetric changes associated with the cathode side of the cathode/SSE interface during repeated charge-discharge cycles lead to the buffer layer's breakage and a subsequent loss of contact [105]. Designing active materials with a large contact area in a porous matrix (with high ionic conductivity) has been a promising strategy to overcome the above challenge and maybe developed using advanced 3-D printing techniques [105].

In order to create a high-quality artificial buffer layer on the SSE surface and/or electrode, numerous synthesis techniques have been reported to be used [105]. These fabrication techniques may be summarized as (i) top-down approach (e.g., magnetron sputtering, spark plasma sintering, etc.) and (ii) bottom-up approach (e.g., sol–gel-derived synthesis, atomic layer deposition, chemical vapor deposition, etc. In recent times, the magnetron sputtering technique has gained commercial success in the synthesis of thin-film SSBs. In contrast, atomic layer deposition has shown promising results in improving the cycle stability of sulfide-type SSBs **[105]. On the other hand,** pulsed laser deposition has been reported to be an attractive method for the preparation of high-quality buffer layers with a stoichiometry, nearly similar to that of the bulk target [105], **This has necessitated the** exploration of the buffer layers through optimization of synthesis techniques. Besides, the expense for large-scale commercial production of thin-film SSBs is also a concern that needs significant attention. Although several characterization techniques have been used to study liquid-based battery systems, a challenge still remains in understanding interfacial behaviour in SSBs [105]. Synthesis of highly reliable artificial buffer layers may be employed to achieve mass production to enable the application of SSEs and help realize the potential of SSBs in electric vehicles.

Although Na-ion battery (SIB) technology, at present, seems to be a replacement for LIBs due to **(i)** lower cost, **(ii)** higher abundance of Na than that if Li, and most interestingly, **(iii)** the chemistry of intercalation **of SIBs** which is very similar to that of LIBs. However, a considerable volume of research needs to be done to make SIB technology keep up with the standards already set by the LIBs, as discussed in an extensive review of SIBs by Palomares et al. [106]. Several Na-based cathodic materials have been investigated, mostly oxides, fluorophosphate, and phosphates [106]. Among the materials investigated to date, oxides do not seem to be a good option due to the complexity of insertion-extraction behaviour [106]. Phosphates, and fluorophosphate, as reported by Palomares et al. [106], may be considered the right choice due to the high stability of these materials and because of the inductive effect of phosphate polyanion produces enormous working potentials. However, the latter needs to be studied in detail to better understand structural characteristics and Na insertion–extraction mechanisms in these materials. This is one area where C-based nanocomposites (primarily, graphite-like materials) may find application as negative electrodes in SIBs. It is primarily due to the presence of $sp^2$ structure that graphene sheets (of heteroatoms) produced during synthesis or post-synthesis process may be expected to benefit the insertion process of Na ions between the stacked layers (present in graphene).

As reported in a recent review on Potassium Ion Batteries (PIBs) [107], the cathode materials used in LIBs may not be suitable for PIBs. For instance, one of the most promising electrode materials available for LIBs is **$LiFePO_4$**, for which the K-analogue has been reported to be practically inactive **due to the demonstration of only** a little activity. Besides, the larger atomic radius of K as compared to that of Li also makes it extremely hard to synthesize $KFePO_4$ with the crystal structure of pure olivine [107].

However, among the investigated cathode materials, the Prussian blue and its analogs have been reported to show excellent electrochemical properties in terms of both specific energy and capacity [107]. Its density is lower than that of layered oxide materials, and limitations include high sensitivity towards water, inadequate specific discharge capacity, etc. [107]. Hence, the significant challenge for designing new cathode materials for PIBs is to have broader and more flexible void space to accommodate high strain produced by the occupation of K ions. In the context of anode materials, the most significant discovery for PIBs is the sandwiching of K into different graphite layers, which has been reported to be advantageous over SIBs since Na cannot be sandwiched effectively into the layers of graphite [107]. The graphite electrode used presently in PIBs has demonstrated excellent specific capacity, rate performance, and cycle life [107]. Although C-based nanocomposites have also been reported as potential candidates to store K ions [105-107], however, they are currently unable to compete with the graphite electrodes in terms of both initial coulombic efficiency and energy density (arising due to the lower operating potential of PIBs).

The more considerable changes (in volume expansion) associated with graphite anode during potentiation needs to be reduced for all practical applications [107]. **Alloyed anodes may be considered as promising candidates for full cell PIBs owing to high-specific capacity; however, certain limitations, which include high operating potential, low initial coulombic efficiency, high volume expansion, etc**. need further investigation. **The materials used for intercalating K into different layers of graphite have been reported to be potential candidates for stable anode materials for practical PIBs [107]; however, detailed investigation in this direction is essential in order to design an anode material with a combination of both high capacity and low operating potential.** Apart from studying individual electrodes in half-cell format, in order to understand the practical performance of a battery, it turns out to be extremely important to fabricate a full-cell [107]. However, at present, only a limited amount of work has been directed in this area. Hence, this turns out to be a field wherein a considerable volume of work may be done, especially in order to increase the initial coulombic efficiency and energy density of C- based nanocomposites so as to be able to completely replace the presently used graphite electrodes in PIBs, both in terms of economy and better electronic properties. This necessitates in-depth investigations for a complete understanding of electrochemistry by finding an appropriate combination of electrodes and electrolytes.

**4.2 From the fundamental viewpoint**

Despite an enormous volume of published literature and reviews on the Electronic properties of Carbon-based nanocomposites, very little is known about the role of different kinds of defects on the electronic properties of these materials, primarily due to lack of experimental information on the direct visualization of these defects. This, of course, becomes highly interesting when the Carbon-based nanocomposite of interest turns out to be crystalline because with the onset of long-range atomic order and the pre-existing short-range order of atoms. It has three entities: (i) the lattice (ii) 1-D dislocations [108] and, most importantly, (ii) different kinds of 2-D interfaces [109], on which the structure-dependent parameters are controlling the electronic properties, or more specifically, the electron transport in the material. The simplest of the interfaces are grain boundaries(GBs) in a polycrystalline material.

Grain boundaries (GBs) act as electron scattering centres and, based on the GB scattering model [77], has also been proposed to study the temperature dependence of electronic conductivity in C-based hard nanocrystalline composite coatings. The segregation of CNTs to GBs in CNT/metal nitride-based nanocomposites is beneficial in the sense that it leads to the enhancement of electronic conductivity in these materials [63]. Although the scattering of electrons at GBs, in the various C- based nanocomposites have now been studied for quite some time using DC resistivity measurements [4]. One must recall some facts viz. dislocations, in any crystalline material, possess a strain field around them and hence act as regions of localized curvature in the microstructure of all crystalline materials [110].

The GBs possess five independent macroscopic degrees of freedom (DOFs) and three dependent microscopic DOFs, which help define the structure of a GB. Based on **these DOFs,** different types of GBs may be distinctly classified, each associated with an entity known as GB energy [14], **in** a polycrystalline material. **Moreover, there is** yet another entity associated with GB segregation known as GB excess [14]. Based on the first fact, it may be expected that there exists a small change in crystallographic orientation of regions adjacent to a dislocation, from which the strain field around the same, maybe calculated based on several orientations coupled with defect imaging microscopy techniques (such as Electron Backscatter Diffraction (EBSD)) (both 2-D and 3-D) for orientation imaging and either TEM and Electron Channeling Contrast Imaging (ECCI) for defect characterization) coupled with the complicated strain tensor analysis [108-110]. The reason as to why the present chapter intends to address these facts is that there exist neither any experimental evidence nor theoretical calculations to account for the influence of GB energy on the electron scattering tendency from the GB. Hence, we claim that the proposed GB scattering model [77] for studying the temperature dependence of electronic conductivity in C- based nanocrystalline composite coatings, also needs to incorporate an additional term accounting for GB energy. Besides, may also be expected to account for GB excess which defines GB segregation [14, 111], for instance in CNT/metal nitride-based nanocomposite materials, thereby making the GB scattering model more robust and versatile for studying the temperature dependence of electronic conductivity in crystalline C-based nanocomposites. One of the main reasons why a study on electron interaction with defects has been missing in the context of C-based nanocomposites is the complexity of understanding crystal structures of these incredibly complex materials and the complexity of sample preparation for experimental investigations. The proposed changes in the GB scattering model [77] describes crystallographic orientation based defect density distribution, and GB segregation in crystalline C-based nanocomposite materials.

Besides, unlike materials for various structural applications, a very limited amount of research has actually been aimed at understanding the "structure-property correlation" for electronic applications in C-based nanocomposites, using the concept of "correlative microscopy" [11-16] which involves correlating the structural information using a number of microscopy techniques from the region of interest in a microstructure with the chemical analysis (upto the atomic level using Atom Probe Tomography (APT) technique [12, 13], for example) from the same using techniques such as Focussed Ion Beaming (FIB) [11, 110-114]. Besides, for the sake of experimental validation, $1^{st}$ principle calculations are also a must. As discussed in the Introductory section of the present chapter, the potential reasons as to why the "correlative microscopy" concept has not been extensively employed in understanding the atomic structure based electronic properties of these materials are the challenges associated with sample preparation for the same and also in characterising a light element as C, which, either in free form or bonded manner, is the most essential constituent in C-based nanocomposites. This, in particular, acts as a potential field of research wherein a lot of future investigations may be carried out to understand electronic properties of C-based nanocomposites based on analysis at the atomic scale. In recent times, the emergence of Artificial Intelligence (AI) and machine learning (ML) guided material design [115] also seems to offer a vast potential in the design of C-based nanocomposites, which may aid material scientists to tailor the electronic properties through microstructural modifications based on AI and ML guided design concepts. This field certainly needs numerous investigations, in the future, as presently, this avenue of research is almost entirely unexplored in C-based nanocomposites.

## 5. Concluding remarks

There remains absolutely no iota of doubt that C-based nanocomposites will find numerous applications in various fields shortly and that the idea of developing these C-based hybrid materials will be utilized to the maximum possible extent, owing to the presence of numerous research possibilities, especially in the field of electronics, both industrially and academically. This will be necessary to render such complex materials as C-based nanocomposites to get closer to being considered more exclusively for day-to-day electronics-related applications in daily life.

**Conflict of Interest**

The authors hereby declare no conflict of interest.

**References:**

1. Zhao Y, Wang LP, Sougrati MT, Feng Z, Leconte Y, Fisher A, Srinivasan M, Xu Z (2017) A Review on Design Strategies for Carbon-Based Metal Oxides and Sulfides

Nanocomposites for High-Performance Li and Na Ion Battery Anodes. Advanced Energy Materials 7:9:1-70.

2. Camargo P H C, Satyanarayana K G, Wypych F (2009) Nanocomposites: Synthesis, Structure, Properties and New Application Opportunities. Materials Research 12:1:1-39,

3. Baibarac M, Romero PG, Cantu M L et al. (2006) Electrosynthesis of the poly (N-vinylcarbazole)/carbon nanotubes composite for applications in the supercapacitor field. Eur Polymer J. 42:2302–2312.

4. Sanjinés R, Abad MD, Vâju Cr, R. Smajda Mionić M, Magrez A (2011) Electrical properties and applications of carbon based nanocomposite materials: An overview. Surface & Coatings Technology 206:727–733.

5. Obreja VVN (2008) On the performance of supercapacitors with electrodes based on carbon nanotubes and carbon activated material-A review. Phys E. 40:2596–2605.

6. Star A, Joshi V, Skarupo S, Thomas D, Gabriel JCP (2006) Gas sensor array based on metal-decorated carbon nanotubes. J Phys Chem B 110:21014.

7. Lu Y, Li J, Han J, Ng HT, Binder C, Partridge C, Meyyapan M (2004) Room temperature methane detection using palladium loaded single-walled carbon nanotube sensors. Chem. Phys Lett 391:344.

8. Ates M, Eker AA, Eker B (2017) Carbon nanotube-based nanocomposites and their applications. J. Adhesion science and Technology. 31:1977–1997.

9. Liu XM, Huang ZD, Oh SW, Zhang B, Ma PC, Yuen MF, Kim JK (2012) Carbon nanotube (CNT)-based composites as electrode material for rechargeable Li-ion batteries: A review. Composites Science and Technology 72:121-144.

10. Wu X, Chen Y, Xing Z, Lam CWK, Pang SS, Zhang W, Ju Z (2019) Advanced Carbon-Based Anodes for Potassium-Ion Batteries. Advanced Energy Materials. 9:1-46.

11. Felfer PJ, Alam T, Ringer SP, Cairney JM (2012) A reproducible method for damage-free site-specific preparation of atom probe tips from interfaces. Microsc Res Techniq 75:484–91.

12. Toji Y, Matsuda H, Herbig M, Choi PP, Raabe D (2014) Atomic-scale analysis of carbon partitioning between martensite and austenite by atom probe tomography and correlative transmission electron microscopy. Acta Mater 65:215–28.

13. Gault B, Moody MP, Cairney JM, Ringer SP (2012) Atom probe crystallography. Mater Today 15:378–86.

14. Herbig M, Raabe D, Li YJ, Choi P, Zaefferer S, Goto S (2014) Atomic-scale quantification of grain boundary segregation in nanocrystalline material. Phys Rev Lett 112: 126103.

15. Singh S, Wanderka N, Murty BS, Glatzel U, Banhart J (2011) Decomposition in multi- component AlCoCrCuFeNi high-entropy alloy. Acta Mater 59: 182–90.

16. Raabe D, Herbig M, Sandlöbes S, Li Y, Tytko D, Kuzmina M, Ponge D, Choi PP (2014) Grain boundary segregation engineering in metallic alloys: A pathway to the design of interfaces. Current opinions in Materials Science 18:253–261.

17. Hornyak GL, Tibbals HF, Dutta J, Moore JJ (2009) Introduction to Nanoscience and Technology. CRC Press, New York.

18. Kuzmany H, Fink J, Mehring M, Roth S (ed) (2000) Electronic Properties of Novel Materials – Molecular Nanostructures. AIP Conference Proceedings 544.

19. Dresselhaus MS, Dresselhaus G, Avouris P (2001) Carbon Nanotubes: Synthesis, Structure, Properties and Applications. 80 Springer, Berlin.

20. Avouris P, Chen Z, Perebeinos V (2007) Carbon-based electronics. Nat. Nanotechnol 2:605.

21. Georgakilas V, Perman JA, Tucek J, Zboril R (2015) Broad Family of Carbon Nanoallotropes: Classification, Chemistry, and Applications of Fullerenes, Carbon Dots, Nanotubes, Graphene, Nanodiamonds, and Combined Superstructures. Chem. Rev., 115: 11:4744–4822

22. Lin Y, Taylor S, Li H, Shiral Fernando KA, Qu L, Wang W (2004) Advances toward bioapplications of carbon nanotubes. J Mater Chem 14:527–541

23. Ajayan PM, Schadler LS, Braun PV (2003) Nanocomposite science and technology. Weinheim (Germany): Wiley-VCH, Verlag GmbH & Co.

24. Nalwa HS (2000) Handbook of nanostructured materials and nanotechnology. 5 Academic Press, New York

25. Modi A, Koratkar N, Lass E, Wei BQ, Ajayan PM (2003) Miniaturized gas ionization sensors using carbon nanotubes. Nature 424:171.

26. Ajayan PM, Lijima S (1992) Smallest carbon nanotube. Nature 358:23.

27. Qiu J, Lia Y, Wang Y, Li W (2004) Production of carbon nanotubes from coal. Fuel Processing Technology 85:1663–1670.

28. Kavita M, Mordina B, Tiwari RK (2016) Thermal and mechanical behaviour of poly(vinyl butyral)- modified novolac epoxy / multi-walled carbon nanotube nanocomposites. J Appl Polym Sci. 133: 43333–43344.

29. Spitalsky Z, Tasis D, Papagelis K, Galiotis C (2010) Carbon nanotube-polymer composites: chemistry, processing, mechanical and electrical properties. Prog Polym Sci. 35: 357–401.

30. Bouchard J, Cayla A, Devaux E, Campagne C (2013) Electrical and thermal conductivities of multi-walled carbon nanotubes-reinforced high performance polymer nanocomposites. Compos Sci Technol.; 86:177–184.

31. Ensafi AA, Soureshjani EH, Asl MJ, Rezaei B (2016) Polyoxometalate-decorated graphene nanosheets and carbon nanotubes, powerful electrocatalysts for hydrogen evolution reaction. Carbon. 99:398–406.

32. Okajima K, Ikeda A, Kamoshita K, Sudoh M (2005) High rate performance of highly dispersed $C_{60}$ on activated carbon capacitor. Electrochim Acta. 51:972–977.

33. Zhou C, Kumar S, Doyle CD, Tour JM (2005) Functionalized single wall carbon nanotubes treated with pyrrole for electrochemical supercapacitor membranes. Chem Mater. 17:1997–2002.

34. Yoon BJ, Jeong SH, Lee KH, Kim HS, Park CG, Han H (2004) Electrical properties of electrical double layer capacitors with integrated carbon nanotube electrodes. Chem Phys Lett. 388:170–174.

35. Koysuren O, Du C, Pan N, Bayram G (2009) Preparation and comparison of two electrodes for supercapacitors: Pani/CNT/Ni and Pani/Alizarin-treated nickel. J Appl Polym Sci. 113:1070–1081.

36. Wang DW, Li F, Zhao J, Ren W, Chen ZG, Tan J, Wu ZS, Gentle I, Lu GQ, Cheng HM (2009) Fabrication of graphene (polyaniline composite paper via in situ anodic electropolymerization for high performance flexible electrode. ACS Nano. 3:1745–1752.

37. Kim JY, Kim KH, Kim KB (2008) Fabrication and electrochemical properties of carbon nanotube/ polypyrrole composite film electrodes with controlled pore size. J Power Sources. 176: 396– 402.

38. An KH, Jeong SY, Hwang HR, Lee YH (2004) Enhanced sensitivity of a gas sensor incorporating single- walled carbon nanotube–polypyrrole nanocomposites. Adv Mater. 16:1005–1009.

39. Cheng G, Zhao J, Tu Y, He P, Fang Y (2005) A sensitive DNA electrochemical biosensor based on magnetite with a glassy carbon electrode modified by multi-walled carbon nanotubes in polypyrrole. Anal Chim Acta. 533:11–16.

40. Limelette P, Schmaltz B, Brault D, Gouineau M, Autret-Lambert C, Roger S, Grimal V, Tran Van (2014) Conductivity scaling and thermoelectric properties of polyaniline hydrochloride. J. Appl. Phys., 115:033712

41. Blaszczyk-Lezak I, Desmaret V, Mijangos C (2016) Electrically conducting polymer nanostructures confined in anodized aluminum oxide templates (AAO). Express Polym Lett. 10: 259–272.

42. Choudhury A, Kar P (2011) Doping effect of carboxylic acid group functionalized multi-walled carbon nanotube on polyaniline. Compos Part B Eng. 42:1641–1647.

43. Wang Y, Zhang S, Deng Y (2016) Semiconductor to metallic behavior transition in multi-wall carbon nanotubes/polyaniline composites with improved thermoelectric properties. Materials Letters. 164:132–135.

44. Taberna PL, Chevallier G, Simon P, Plée D, Aubert T (2006) Activated carbon-carbon nanotube composite porous film for supercapacitor applications. Mater Res Bull. 41:478–484.

45. Navarro-Flores E, Omanovic S (2005) Hydrogen evolution on nickel incorporated in three- dimensional conducting polymer layers. J Mol Catal A Chem. 242:182–194.

46. Huq MM, Hsieh CT, Ho CY (2016) Preparation of carbon nanotube-activated carbon hybrid electrodes by electrophoretic deposition for supercapacitor applications. Diamond Related Mater. 62:58–64.

47. Khomenko V, Raymundo-Pinero E, Beguin F (2008) High-energy density graphite/AC capacitor in organic electrolyte. J Power Sources. 177:643–651.

48. Qiu J, Wu X, Qiu T (2016) High electromagnetic wave absorbing performance of activated hollow carbon fibers decorated with CNTs and Ni nanoparticles. Ceram Int. 42:5278–5285.

49. Gangupomu RH, Sattler ML, Ramirez D (2016) Comparative study of carbon nanotubes and granular activated carbon: physicochemical properties and adsorption capacities. J Hazard Mater. 302:362–374.

50. Rudge A, Davey J, Raistrick I, Gottesfeld S (1994) Conducting polymers as active materials in electrochemical capacitors. J. Power Sources 47:89.

51. Bouchard J, Cayla A, Odent S, Lutz V, Devaux E, Campagne C (2016) Processing and characterization of polyethersulfone wet-spun nanocomposite fibres containing multi walled carbon Nanotubes. Synthetic Metals. 217:304-313.

52. Yuan D, Yang W, Ni J, Gao L (2015) Sandwich structured MoO2@TiO2@CNT nanocomposites with high-rate performance for lithium ion batteries. Electrochim Acta. 163:57–63.

53. Wang YG, Wang ZD, Xia YY (2005) An asymmetric supercapacitor using RuO2/TiO2 nanotube composite and activated carbon electrodes. Electrochim Acta. 50:5641–5646.

54. Alam RS, Moradi M, Nikmanesh H (2016) Influence of multi-walled carbon nanotubes (MWCNTs) volume percentage on the magnetic and microwave absorbing properties of $BaMg_{0.5}Co_{0.5}TiFe_{10}O_{19}$/MWCNTs nanocomposites. Mater Res Bull. 73:261–267.

55. Yuan QH, Zeng XS, Liu Y, Luo L, Wu J, Wang Y, Zhou G (2016) Microstructure and mechanical properties of AZ91 alloy reinforced by carbon nanotubes coated with MgO. Carbon. 96:843–855.

56. Islam MS, Deng Y, Tong L, Roy AK, Minett AI, Gomes VG (2016) Grafting carbon nanotubes directly onto carbon fibers for superior mechanical stability: towards next generation aerospace composites and energy storage applications. Carbon. 96:701–710.

57. Wu G, Ma L, Liu L, Wang Y, Xie F, Zhong Z, Zhao M, Jiang B, Huang Y (2016) Interface enhancement of carbon fiber reinforced methylphenylsilicone resin composites modified with silanized carbon nanotubes. Mater Des. 89:1343–1349.

58. Wang Y, Colas G, Filleter T (2016) Improvements in the mechanical properties of carbon nanotube fibers through graphene oxide interlocking. Carbon. 98:291–299.

59. Yang LJ, Cui JL, Wang Y, et al (2016) Research progress on the interconnection of carbon nanotubes. New Carbon Mater. 31:1–17.

60. Tamrakar S, An Q, Thostenson ET, Rider AN, Haque BZ (Gama), Gillespie JW Jr (2016) Tailoring interfacial properties by controlling carbon nanotube coating thickness on glass fibers using electrophoretic deposition. ACS Appl Mater Interfaces. 8:1501–1510.

61. Flahaut E, Peigney A, Laurent Ch, Ch. Chastel MF, Rousset A (2000) Acta Mater., 48:3803.

62. Kymakis E, Alexandou I Amaratunga GAJ (2002) Single-walled carbon nanotube–polymer composites: electrical, optical and structural investigation. Synth. Met., 127:59.

63. Jiang L, Gao L (2005) Carbon nanotubes–metal nitride composites: a new class of nanocomposites with enhanced electrical properties. J. Mater. Chem. 15:260–266.

64. Peigney A, Laurent Ch Rousset A (1997) Key Eng. Mater. 743:132–136.

65. Rao CNR, Satishkumar BC, Govindaraj A, Nath M (2001) Nanotubes. Chem Phys Chem. 2:78.

66. Sun Z, Zhang J, Yin L, Hu G, Fang R, Cheng HM, Li F (2017) Conductive porous vanadium nitride/graphene composite as chemical anchor of polysulfides for lithium-sulfur batteries. Nat Comm; 8: 14627.

67. Zhao JG, Yang LX, Li FY, Yu RC, Jin CQ (2008) Electrical property evolution in the graphitization process of activated carbon by high-pressure sintering. Solid State Sci. 10:1947

68. Staryga E, Bak GW (2005) Relation between physical structure and electrical properties of diamond-like carbon thin films. Diamond Relat. Mater. 14:23.

69. Mott NF, Davis EA (1979) Electronic Processes in Non-Crystalline Materials. Clarendon Press, Oxford.

70. Shimikawa K, Miyake K (1989) Hopping transport of localized $\pi$ electrons in amorphous carbon films. Phys. Rev. B 39:7578.

71. Godet C, Kleider JP, Gudovskikh AS (2007) Frequency scaling of AC hopping transport in amorphous carbon nitride. Diamond Relat. Mater. 16:1799.

72. Vishwakarma PN, Subramanyam SV (2006) Hopping conduction in boron-doped amorphous carbon films. J. Appl. Phys. 100:113702.

73. Zhong DH, Sano H, Uchiyama Y, Kobayashi K (2000) Effect of low-level boron doping on the oxidation behavior of polyimide-derived carbon films. Carbon 38:1199.

74. Sikora A, Berkesse A, Bourgeois O, Garden JL, Guerret-Piécourt C, Rouzaud JN, Loir AS, Garrelie F, Donnet C (2009) Structural and electrical characterization of boron-containing diamond-like carbon films deposited by femtosecond pulsed laser ablation. Solid State Sci. 11:1738.

75. Xue B, Chen P, Hong Q, Lin J, Tan KL (2001) Growth of Pd, Pt, Ag and Au nanoparticles on carbon nanotubes. J. Mater. Chem. 11:2378.

76. Meiners T, Frolov T, Rudd RE, Dehm G, Liebscher CH (2020) Observations of grain-boundary phase transformations in an elemental metal. Nature 579:375–378.

77. Mishnaevsky Jr LL (2007) Computational Mesomechanics of Composites. John Wiley England.

78. Nigro A, Nobile G, Rubino MG, Vaglio R (1988) Electrical resistivity of polycrystalline niobium nitride films. Phys. Rev. B 37:3970.

79. Yu Z, Tetard L, Zhai L, Thomas J (2013) Supercapacitor electrode-materials nanostructures from 0 to 3 dimensions. Energy & Environ Sci. 8:702–730.

80. Borenstien A, Noked M, Okashy S, Aurbach D (2013) Composite carbon nanotubes (CNT)/activated carbon electrodes for non-aqueous supercapacitors using organic electrolyte solutions. J Electrochem Soc. 160:A1282–A1285.

81. Davies A, Yu A (2011) Material advancements in supercapacitors: from activated carbon to carbon nanotube and graphene. Can J Chem Eng. 89:1342–1357.

82. Stankovich S, Dikin DA, Dommett GHB, Kohlhaas KM, Zimney EJ, Stach EA, Piner RD, Nguyen ST, Ruoff RS (2006) Graphene-based composite materials. Nature 442:282–6.

83. Geim AK, Novoselov KS (2007) The rise of graphene. Nat Mater 6:183–191.

84. Geng Y, Wang SJ, Kim JK (2009) Preparation of graphite nanoplatelets and graphene Sheets. J Colloid Interface Sci 336:592–8.

85. Zheng QB, Ip WH, Lin XY, Yousefi N, Yeung KK, Li Z, Kim JK (2011) Transparent conductive films consisting of ultralarge graphene sheets produced by Langmuir–Blodgett assembly. ACS Nano 5:6039–6051.

86. Lian PC, Zhu XF, Liang SZ, Li Z, Yang WS, Wang HH (2010) Large reversible capacity of high quality graphene sheets as an anode material for lithium-ion batteries. Electrochim Acta 55:3909–3914.

87. Wu ZS, Ren WC, Wen L, Gao LB, Zhao JP, Chen ZP (2010) Graphene anchored with Co3O4 nanoparticles as anode of lithium ion batteries with enhanced reversible capacity and cyclic performance. ACS Nano 4:3187–3194.

88. Su FY, You CH, He YB, Lv W, Cui W, Jin FM, Li B, Yang QH, Kang F (2010) Flexible and planar graphene conductive additives for lithium-ion batteries. J Mater Chem 20:9644–9650.

89. Guo Y, Wang T, Chen F, Sun X, Li X, Yu Z, Wan P, Chen X (2016) Hierarchical graphene–polyaniline nanocomposite films for high-performance flexible electronic gas sensors. Nanoscale, 8:23:12073–12080.

90. Chang H, Wu H (2013) Graphene-based nanocomposites: preparation, functionalization, and energy and environmental applications. Energy & Environmental Science 6(12): 3483.

91. Geim AK (2009) Graphene: status and prospects. Science 324:1530-1534.

92. Allen MJ, Tung VC, Kaner RB (2009) Honeycomb carbon: a review of graphene. Chem. Rev. 110:132-145.

93. Rao CNR, Sood AK, Subrahmanyam KS, Govindaraj A (2009) Graphene: the new two-dimensional nanomaterial. Angew. Chem. Int. Edit. 48:7752-7777.

94. Chang HX Wu HK (2013) Graphene-Based Nanomaterials: Synthesis, Properties, and Optical and Optoelectronic Applications; Adv. Funct. Mater. 23:1984-1997.

95. Yang K, Feng L, Shi X, Liu Z (2013) Nano-graphene in biomedicine: theranostic applications. Chem. Soc. Rev. 42:530-547.

96. Zhu Y, Murali S, Cai W, Li X, J. Suk W, Potts JR, Ruoff RS (2010) Graphene and graphene oxide: synthesis, properties, and applications. Adv. Mater. 22:3906-3924

97. Du X, Skachko I, Barker A, Andrei EY (2008) Approaching ballistic transport in suspended graphene. Nat. Nanotechnol. 3:491-495.

98. Huang X, Yin ZY, Wu SX, Qi XY, He QY, Zhang QC, Yan QY, Boey F, Zhang H (2011) Graphene-based materials: synthesis, characterization, properties, and applications. Small 7:1876-1902.

99. Lee C, Wei XD, Kysar JW Hone J (2008) Measurement of the elastic properties and intrinsic strength of monolayer graphene. Science 321:385-388.

100. Balandin AA, Ghosh S, Bao W, Calizo I, Teweldebrhan D, Miao F, Lau CN (2008) Superior thermal conductivity of single-layer graphene. Nano Lett. 8:902-907.

101. Nair R, Blake P, Grigorenko A, Novoselov K, Booth T, Stauber T, Peres N, Geim A (2008) Fine structure constant defines visual transparency of graphene. Science 320: 1308.

102. Lin YM, Dimitrakopoulos C, Jenkins KA, Farmer DB, Chiu HY, Grill A, Avouris P (2010) 100-GHz transistors from wafer-scale epitaxial graphene. Science 327:662-662.

103. Liao L, Lin Y-C, Bao M, Cheng R, Bai J, Liu Y, Qu Y, Wang KL, Huang Y Duan X (2010) High-speed graphene transistors with a self-aligned nanowire gate. Nature 467: 305-308.

104. Schwierz F (2010) Graphene transistors. Nat. Nanotechnol. 5:487-496.

105. Du M, Liao K, Lu Q, Shao Z (2019) Recent advances in the interface engineering of solid-state Li-ion batteries with artificial buffer layers: Challenges, materials, construction, and characterization. Energy and Environmental Science 12(6):1780-1804.

106. Palomares V, Serras P, Villaluenga I, Hueso KB, Carretero-Gonzalez J, Rojo T (2012) Na-ion batteries, recent advances and present challenges to become low cost energy storage systems. Energy Environ. Sci. 5:5884.

107. Rajagopalan R, Tang Y, Ji X, Jia C, Wang H (2020) Advancements and Challenges in Potassium Ion Batteries: A Comprehensive Review. Adv. Funct. Mater. 1909486.

108. Cottrell AH (1949) Theory of dislocations. B. Chalmers (Ed.), Progress in Metal Physics Chapter. II, p. 1-52.

109. Gleiter H (1983) On the Structure of Grain Boundaries in Metals. In: Latanision R.M., Pickens J.R. (eds) Atomistics of Fracture. Springer Boston MA

110. Zaafarani N, Raabe D, Roters F, Zaefferer S (2008) On the origin of deformation-induced rotation patterns below nanoindents. Acta Mater. 56(1):31-42.

111. Zaafarani N, Raabe D, Singh RN, Roters F, Zaefferer S (2006) Three-dimensional investigation of the texture and microstructure below a nanoindent in a Cu single crystal using 3D EBSD and crystal plasticity finite element simulations. Acta Mater. 54:1863–1876.

112. Wheeler J, Mariani E, Piazolo S, Prior DJ, Trimby PJ, Drury MR (2009) The weighted Burgers vector: a new quantity for constraining dislocation densities and types using electron backscatter diffraction on 2D sections through crystalline materials. J Microscopy 233: 482–494.

113. Gutierrez-Urrutia I, Zaefferer S, Raabe D (2013) Coupling of Electron Channeling with EBSD: Toward the Quantitative Characterization of Deformation Structures in the SEM. JOM, 65( 9):1229-1236.

114. Stoffers A, Cojocaru-Mirédin O, Seifert W, Zaefferer S, Riepe S, Raabe D (2015) Grain boundary segregation in multicrystalline silicon: correlative characterization by EBSD, EBIC, and atom probe tomography. Prog. Photovolt: Res. Appl. 23: 1742–1753.

115. Huber L, Hadian R, Grabowski B, Neugebauer J (2018) A machine learning approach to model solute grain boundary segregation. npj Computational Materials 64(1):1-8.